\documentclass{phb}
\usepackage{graphicx}

\begin{document}

\begin{frontmatter}

\title{Plane dimpling and Cu 4$s$\/ hybridization in YBa$_2$Cu$_3$O$_x$}

\author[address]{J. R\"ohler\thanksref{thank}}

\address[address]{Universit\"at zu K{\"o}ln,
Z\"ulpicher Str. 77, D-50937 K{\"o}ln, Germany}

\thanks[thank]{E-mail: abb12@uni-koeln.de}

\begin{abstract}

Oxygen doping dimples the CuO$_2$ planes of
YBa$_{2}$Cu$_{3}$O$_{6.8-6.92}$ by displacing copper normal to the
planes and further towards Ba.  The correlated oxygen displacements,
however, are constrained to the in-plane axes.  This displacement
pattern is discussed in terms of doping dependent Cu 4$s$--3$d$
hybridizations.
\end{abstract}

\begin{keyword}
Electronic structure, superconductivity, YBa$_2$Cu$_3$O$_7$, EXAFS
\end{keyword}
\end{frontmatter}


The atomic positions of the planar copper and oxygen atoms in the
high-$T_c$ cuprates exhibit important deviations from the ideal square
planar geometry.  Most significant are the static displacements normal
to the 2D translational axes, the so-called plane buckling or
plane dimpling.  We discern the LTT-, LTO-type tilts of the $\it rigid$\/
CuO$_6$ octahedra in the La 214 family, and the incommensurately
modulated tilts of $\it rigid$\/ CuO$_5$ pyramids in Bi 2212, from the
$\it soft$  CuO$_4$ squares allowing for ``dimples'' in e.g.
YBa$_{2}$Cu$_{3}$O$_{x}$.

In this short note we address a significant correlation of displacive
degrees of freedom in the CuO$_2$ planes of YBa$_{2}$Cu$_{3}$O$_x$
upon doping from $6.806<x<6.92=x_{opt}$.  The CuO$_2$ planes of
YBa$_{2}$Cu$_{3}$O$_x$ are best described as stacks of O2,3 and Cu2
layers with an interlayer distance of $s\simeq 0.24$ \AA. The O2,3
layers are located closest to the Y layers, and the Cu2 layers closest
to the Ba layers.  Y-EXAFS measurements of YBa$_{2}$Cu$_{3}$O$_{x}$ as
a function of doping \cite{Roe} have shown, that doping pulls down the
Cu2 atoms towards the Ba layer whereas the spacings between the O2,3
and Y layers remain unaffected.  Hence the O2,3--Cu2 interlayer
distance increases, in the metallic and superconducting regime
relatively up to $\delta s \simeq 0.05$ {\AA} \cite{Roe,Dev}.  It is
important to note that the O2,3--Y interlayer distances do not alter.
Thus the reservoir layer accomodates the contraction of the $c$\/-axis
parameter.  From our doping dependent Y-EXAFS the displacements of Cu2
atoms along the $c$\/-axis turn out to occur at fixed
$R_{Y-Cu2}=3.202$(5) {\AA} $= const$.  Phenomenologically this
constraint couples the perpendicular displacements of Cu2 (dimples)
with the in-plane displacements of the cation sublattices and, most
important, with those of the planar oxygens O2,3.  Fig.  1 exhibits
schematically these correlated displacements by comparing two subcells
with differently dimpled CuO$_2$ planes (exaggerated scale).  Doping
is shown to pull the Cu2 towards the Ba layer and to push thereby the
O2,3 horizontally along the dashed line.  The latter indicates the
projected average in-plane translational axes of the O2,3.  Different
orthorhombicities, $(b-a)/(b+a)$, are clearly visible in Fig.  1
$(top)$.\/ These correspond to the doping dependencies of the
$a,b$\/-axes close to $x_{opt}$: $\partial a/\partial x < 0, \partial
b/\partial x > 0$, and $\mid\partial a/\partial x\mid >\\\mid\partial
b/\partial x\mid$.

\begin{figure}[btp]
\begin{center}\leavevmode
\includegraphics[width=0.86\linewidth]{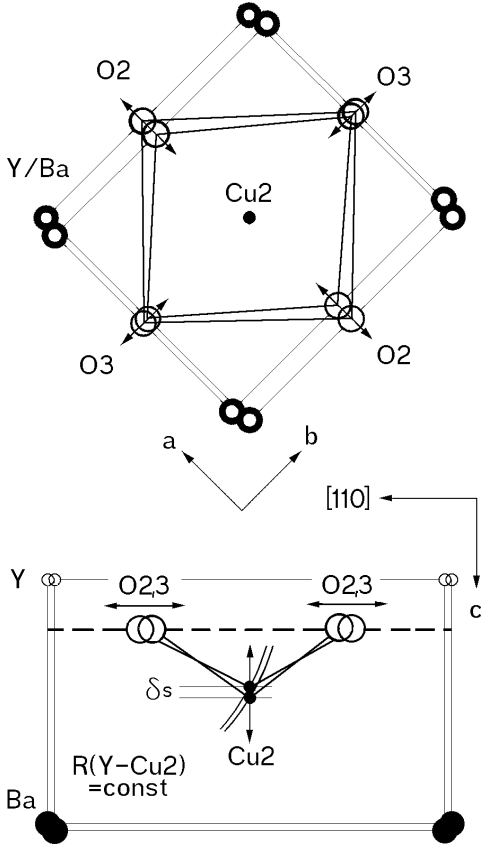}
\caption{
Correlated displacements of the planar Cu2 and O2,3 atoms at different
oxygen dopings. See text.
}\label{fig4}\end{center}\end{figure}

The planar dimples are shown by Andersen {\it et al.} \cite{And} to
have important effects on the single electron structure of
YBa$_{2}$Cu$_{3}$O$_{7}$ and its Fermi surface, in particular on the
dispersion of the dominant Cu $3d _{x^2-y^2}$--O$2p_{x,y}$
$dp\sigma$\/-band close to ($\pi,0$), and (0,$\pi$).  Or the other way
round, the single electron structure of the cuprates generically
activates atomic displacements perpendicular to the planes affecting
thereby also the Fermi surface.  In fact hybridizations of the Cu $4s$
orbital are at heart of all important degrees of freedom normal to the
CuO$_2$ planes.  Cu $4s$ belongs to the minimum set of orbitals
describing a realistic single electron structure, and may not be
integrated out as the other high energy orbitals.  (Cu 3d$_{z^2}$
hybridization turns out to be a negligible quantum chemical parameter
of the electronic structure, both, experimentally and theoretically).
Plane dimpling is found to be activated by remote hybridization of the
Cu $4s$ with the Cu 3$d_{x^2-y^2}$ orbitals.  It works via the Cu
3$d_{xz}$, Cu 3$d_{yz}$ and their overlap with the O 2p$_{x,y}$,
overlapping with $3d_{x^2-y^2}$.  As a result perfectly flat CuO$_{2}$
planes get intrinsically instable upon doping, but may be stabilized
by atomic displacements along the normal axis.

Summarizing we emphazise that doping of the CuO$_2$ planes in
YBa$_{2}$Cu$_{3}$O$_{7}$ activates the Cu2 atoms to be displaced
normal to the plane, and the planar oxygens to be slaved displacing
along the in-plane translational axes.  The Cu $4s$--$3d$
hybridization seems to be the crucial quantum chemical parameter
controling this correlation of in- and out-of-plane displacements, and
possibly also the related interplanar and intraplanar electronic
degrees of freedom.


.


\begin{thebibliography}{9}

\bibitem{Roe} J. R\"ohler {\it et al.}, in {\it Work\-shop on
High\--$T_c$ Superconduc-\\tivity 1996: Ten Years after the
Discovery}, E. Kaldis, E. Liarokapis, K. A. M\"uller (eds.), NATO ASI
Series {\bf E343} (Kluwer, Dordrecht, 1997), 469; J. R\"ohler {\it et al.},
Physica {\bf C 282-287}, 182 (1997).

\bibitem{Dev} The insulating ``background'' dimpling of 0.2 {\AA } has
to substracted.  It is most likely due to the un\-screened charge
contrast between the Y$^{3+}$ and Ba$^{2+}$ layers, acting differently
on the Cu$^{2+}$ and O$^{2-}$ ions as discussed by Deveraux {\it et
al.}, cond-mat/9712128.  Electrostatic polarization, however, cannot
cause the observed increase of the dimpling at the insulator-metal
transition and beyond.  The onset of metallic screening in the planes
is expected to reduce the dimpling instead of enhancing it.  Therefore
the electrostatic mechanism is not at the origin of the modified
$B_{1g}$ Raman modes, either.  The {\it sd} hybridization effects are
more likely.

\bibitem{And} O. K. Andersen {\it et al.}, J. Phys.  Chem.  Solids {\bf
12}, 1573 (1995); O. K. Andersen {\it et al.}, J. Low Temp. Phys.
{\bf 105}, 286 (1996).

\end{thebibliography}
\end{document}